\documentclass{jpsj3}

\title{The Effect of $f$-$d$ Magnetic Coupling in Multiferroic $R$MnO$_3$ Crystals}

\author{Masaaki \textsc{Hitomi}, Mizuaki \textsc{Ehara}, Mitsuru \textsc{Akaki}, Daisuke \textsc{Akahoshi}, \newline
and Hideki \textsc{Kuwahara}\thanks{E-mail: h-kuwaha@sophia.ac.jp}}

\inst{Department of Physics, Sophia University, Tokyo 102-8554}

\abst{We have established detailed magnetoelectric phase diagrams of (Eu$_{0.595}$Y$_{0.405}$)$_{1-x}$Tb$_x$MnO$_3$ ($0 \le x \le 1$) and (Eu,Y)$_{1-x}$Gd$_x$MnO$_3$ ($0 \le x \le 0.69$), whose average ionic radii of $R$-site ($R$: rare earth) cations are equal to that of Tb$^{3+}$, in order to reveal the effect of rare earth 4$f$ magnetic moments on the magnetoelectric properties. 
In spite of the same $R$-site ionic radii, the magnetoelectric properties of the two systems are remarkably different from each other.
A small amount of Tb substitution on $R$ sites ($x \sim 0.2$) totally destroys ferroelectric polarization along the $a$ axis ($P_a$), and an increase in Tb concentration stabilizes the $P_c$ phase. 
On the other hand, Gd substitution ($x \sim 0.2$) extinguishes the $P_c$ phase, and slightly suppresses the $P_a$ phase. 
These results demonstrate that the magnetoelectric properties of $R$MnO$_3$ strongly depend on the characteristics of the rare earth 4$f$ moments.}

\kword{multiferroic, magnetoelectric property, perovskite manganese oxide, phase diagram, 4$f$ moment}

\begin{document}
\maketitle
\newpage
\section{Introduction}

In recent years, a new class of multiferroics, which have both magnetic and ferroelectric orders, has been attracting increasing interest. $R$MnO$_3$ ($R$: rare earth) of orthorhombic perovskite structure is one of the most typical examples. In $R$~=~La~-~Eu, $R$MnO$_3$ has the $A$-type antiferromagnetic ground state of Mn 3$d$ spins.\cite{Goto} 
As the perovskite $R$-site ionic radius decreases, the magnetic ground state changes from $A$-type antiferromagnetic order to noncollinear transverse spiral antiferromagnetic order around $R$~=~Gd, owing to magnetic frustration between the nearest neighbor (NN) ferromagnetic and next NN antiferromagnetic interactions among Mn spins.\cite{prbkim,Kenzelmann,arima,katsura,kaji} In $R$MnO$_3$ where $R$~=~Tb~or~Dy, noncollinear transverse spiral antiferromagnetic order breaks the inversion symmetry through the inverse Dzyaloshinskii-Moriya (DM) interaction, resulting in the appearance of ferroelectricity.\cite{Kenzelmann,katsura,most,arimaPRB} The spin helicity or vector spin chirality of the spiral spin structure can be changed by external magnetic fields, then the direction of the ferroelectric polarization due to the spin chirality can also be controlled by the fields.\cite{nature} 
As the $R$-site ionic radius decreases from $R$~=~Dy to $R$~=~Ho~-~Lu,~and~Y, the ground state changes from noncollinear spiral antiferromagnetic to collinear $E$-type antiferromagnetic order.\cite{Ivan,Goto} $R$MnO$_3$ with $E$-type antiferromagnetic order also exhibits ferroelectric behavior that is caused not by the inverse DM interaction but by inverse Goodenough-Kanamori interaction, i.e., the symmetric exchange striction.\cite{Ivan,iGK} 

The magnetoelectricity of $R$MnO$_3$ is largely affected not just by the magnetic frustration of Mn 3$d$ spins but also by 4$f$ magnetic moments of $R^{3+}$.\cite{GdTb,japnoda,MRSkuwa,hem,EuY045,EuY} 
In our previous works,\cite{mmmnoda} to clarify the effect of 4$f$ moments, we made a comparative study of TbMnO$_3$, Y$_{0.31}$Gd$_{0.69}$MnO$_3$, and Eu$_{0.595}$Y$_{0.405}$MnO$_3$, whose average ionic radii of $R$ sites are equal to that of Tb$^{3+}$. Although they have the same ionic radii of $R$-site cations, the magnetoelectric properties of these three compounds are quite different. In 4$f$-moment-free Eu$_{0.595}$Y$_{0.405}$MnO$_3$,\cite{japnoda,kaji2} ferroelectric polarization along the $c$-axis ($P_c$) emerges below $T_{{\rm FE}c}$~=~25~K; $P_c$ disappears at $T_{{\rm FE}a}$~=~23~K\@, below which ferroelectric polarization is observed parallel to the $a$ axis ($P_a$) [Fig.~\ref{fig1}(a)]. In TbMnO$_3$, the polarization flop does not occur without a magnetic field, and the polarization of the ferroelectric ground state is along the $c$-axis.\cite{prbkim,arimaPRB,nature}
The difference between their magnetoelectric properties is attributed to the presence of Tb$^{3+}$ 4$f$ moments. Tb$^{3+}$ has an anisotropic 4$f$ moment, because of its large orbital angular momentum ($L = 3$). The ordered Tb$^{3+}$ moments below 7~K are along either of two Ising axes within the $ab$ plane at 57$^\circ$ off the $b$-axis.\cite{Tb57} Thus, in TbMnO$_3$, magnetic coupling between 4$f$ Ising-like moments and Mn 3$d$ spins has an effect similar to that of applying external magnetic fields to the Mn$^{3+}$ sublattice. 
Hereafter, we refer to these effective magnetic fields as ``internal magnetic fields".
The 4$f$-moment effect of Gd$^{3+}$ is opposite to that of Tb$^{3+}$. In Y$_{0.31}$Gd$_{0.69}$MnO$_3$, only the $P_a$ phase is found in a zero magnetic field.\cite{MRSkuwa,mmmnoda} The 4$f$ moment of Gd$^{3+}$ is of the Heisenberg type without anisotropy, because $L = 0$. Consequently, the Mn$^{3+}$ sublattice of Y$_{0.31}$Gd$_{0.69}$MnO$_3$ does not feel the internal magnetic field along the $a$-axis, which develops $P_c$. 

As mentioned above, the rare earth 4$f$ moment has a significant effect on the magnetoelectric properties of $R$MnO$_3$, but the effect is not well understood yet. 
In this study, to reveal the contribution of 4$f$ moments to the magnetoelectric properties of $R$MnO$_3$, we have established detailed magnetoelectric phase diagrams of (Eu$_{0.595}$Y$_{0.405}$)$_{1-x}$Tb$_x$MnO$_3$ and (Eu,Y)$_{1-x}$Gd$_x$MnO$_3$, whose average $R$-site ionic radii are fixed at that of Tb$^{3+}$.

\section{Experiment}
We prepared a series of (Eu$_{0.595}$Y$_{0.405}$)$_{1-x}$Tb$_x$MnO$_3$ ($0\leq x\leq 1$) and (Eu,Y)$_{1-x}$Gd$_x$MnO$_3$ ($0\leq x\leq 0.69$) crystals with several substitution levels, where the average ionic radii of $R$-site are equal to that of Tb$^{3+}$. In the latter system, therefore, the ratio of Eu to Y is tuned appropriately depending on Gd concentration~$x$. It should be noted that $x$~=~0.69 is the limit of Gd concentration for retaining the average $R$-site ionic radii at that of Tb$^{3+}$\@. The average ionic radii were calculated on the basis of the Shannon's ionic radius table.\cite{acta} The starting materials, Eu$_2$O$_3$, Y$_2$O$_3$, Tb$_4$O$_7$, Gd$_2$O$_3$, and Mn$_3$O$_4$ were weighted to a prescribed ratio and mixed by using ethanol. The mixture was heated at 1073~K in air with a few intermediate grindings, and sintered at 1473~K in air. The resultant powder was formed to a cylindrical shape (6~mm~$\phi$~$\times$~100~mm) with use of a hydrostatic pressure of $300\sim350$~kgf/cm$^2$ to make a feed rod. Single crystalline samples were grown under Ar atmosphere ($\sim$~3~atm) at a feed rate of 3~$\sim$~6~mm/h by a floating zone method. We performed X-ray diffraction (XRD) measurements on the obtained crystals at room temperature. The Rietveld analysis of powder XRD data revealed that all the samples have an orthorhombic $Pbnm$ structure without any impurity phases or any phase segregation. All the single crystalline samples used in this study were cut along the crystallographic principal axes into a rectangular shape using an X-ray back-reflection Laue technique. The dielectric constant and the ferroelectric polarization in magnetic fields were measured using a temperature-controllable cryostat equipped with a superconducting magnet that provides a magnetic field up to 8~T\@. The dielectric constant measurement was performed with an $LCR$ meter at a frequency of 10~kHz (Agilent,~4284A). 
The pyroelectric current to obtain the spontaneous electric polarization was measured in a warming process at a rate of 4~K/min after the samples were cooled from 60~to~5~K in a poling electric field of 300~$\sim$~500~kV/m.  
The magnetic properties were measured using a commercial apparatus (Quantum~Design,~PPMS-9T).

\section{Results}
\begin{figure*}[tb]
\begin{center}
\includegraphics[width=1 \textwidth, clip]{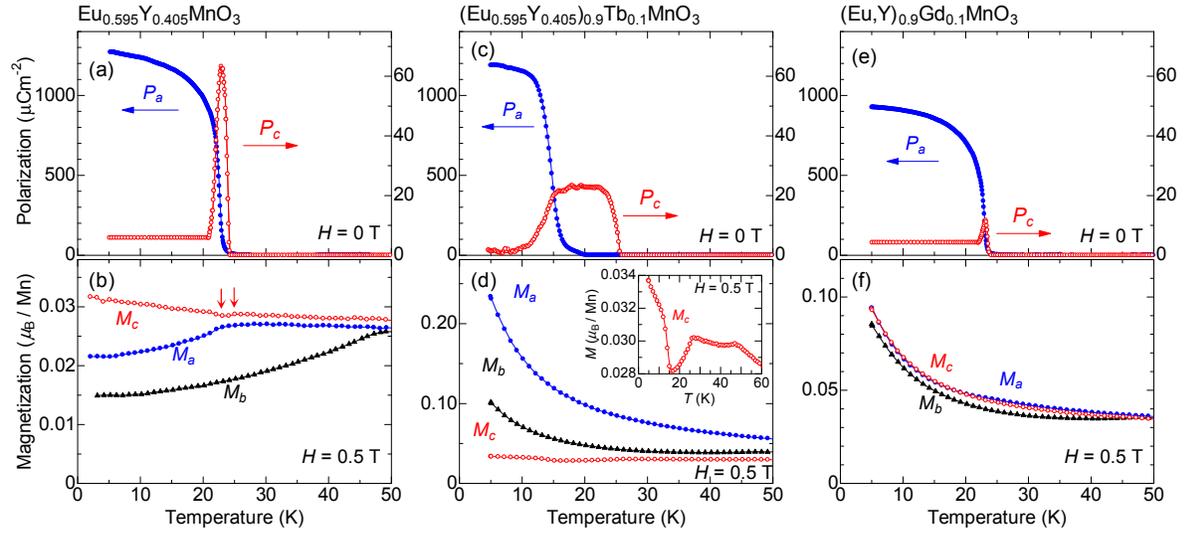}
\end{center}
\caption{(Color online) Temperature dependence of the ferroelectric polarization along the $a$ ($P_a$) and $c$ ($P_c$) axes (upper) and magnetization along the $a$ ($M_a$), $b$ ($M_b$), and $c$ ($M_c$) axes (lower) of Eu$_{0.595}$Y$_{0.405}$MnO$_3$ [(a)~and~(b)],\cite{japnoda} (Eu$_{0.595}$Y$_{0.405}$)$_{0.9}$Tb$_{0.1}$MnO$_3$ [(c) and (d)], and (Eu,Y)$_{0.9}$Gd$_{0.1}$MnO$_3$ [(e) and (f)]. Arrows in (b) show ferromagnetic transition temperatures of Eu$_{0.595}$Y$_{0.405}$MnO$_3$. Inset of (d) shows magnification of $M_c$ in (Eu$_{0.595}$Y$_{0.405}$)$_{0.9}$Tb$_{0.1}$MnO$_3$. Applied magnetic field of $H = 0.5$~T was directed along each principal axes.}
\label{fig1}
\end{figure*}

Let us start with the 4$f$-moment-free compound, Eu$_{0.595}$Y$_{0.405}$MnO$_3$.\cite{japnoda,EuY045} 
Figure \ref{fig1} shows the temperature dependence of (a) the ferroelectric polarization in a zero magnetic field and (b) the magnetization in $H$~=~0.5~T. $P_c$ emerges below $T_{{\rm FE}c}$~=~25~K\@, and the magnetization parallel to the $c$-axis ($M_c$) shows a slight decrease at the same time. However, no anomaly is clearly discerned in the magnetization parallel to the $a$-axis ($M_a$) or the $b$-axis ($M_b$) at $T_{{\rm FE}c}$. 
The $P_c$ phase is replaced by $P_a$. That is, the direction of the ferroelectric polarization changes from the $c$-axis to the $a$-axis at $T_{{\rm FE}a}$. 
This polarization flop is caused by the magnetic phase transition from $bc$ spiral ($P_c$) to $ab$ spiral ($P_a$) antiferromagnetic order, which accompanies a decrease in $M_a$ and an increase in $M_c$. 
The magnitude of $P_a$ is much larger than that of $P_c$. This is because the polarization flop from the $c$-axis to the $a$-axis immediately below $T_{{\rm FE}c}$ before the saturation of $P_c$.  
The decrease in $M_b$ below 47~K is due to an incommensurate antiferromagnetic ordering with a collinear sinusoidal modulation along the $b$-axis, which is considered unrelated to the magnetoelectric properties. 

\begin{figure}[tb]
\begin{center}
\includegraphics[width=0.4 \textwidth, clip]{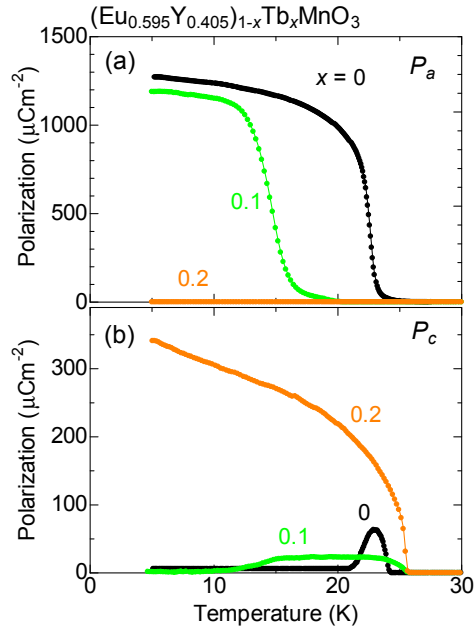}
\end{center}
\caption{(Color online) Temperature dependence of the ferroelectric polarization along the $a$ ($P_a$,~upper) and $c$ ($P_c$,~lower) axes of (Eu$_{0.595}$Y$_{0.405}$)$_{1-x}$Tb$_x$MnO$_3$ with 0~$\le x \le$~0.2 in~a~zero magnetic~field.}
\label{fig2}
\end{figure}

Tb doping into Eu$_{0.595}$Y$_{0.405}$MnO$_3$ greatly affects the magnetoelectric properties. 
In (Eu$_{0.595}$Y$_{0.405}$)$_{0.9}$Tb$_{0.1}$MnO$_3$\@, $T_{{\rm FE}a}$ shifts to lower temperatures by 8~K\@, and consequently the $P_c$ region is broadened compared to that of Eu$_{0.595}$Y$_{0.405}$MnO$_3$ [Fig.~\ref{fig1}(c)]. $M_c$ shows a clear decrease at $T_{{\rm FE}c}$~(=~25~K), and then increases below $T_{{\rm FE}a}$~(=~15~K) [inset~of~Fig.~\ref{fig1}(d)]. These changes are also consistent with the magnetic transition from the $bc$ spiral ($P_c$) to the $ab$ spiral ($P_a$). Note that the magnetic anisotropy of (Eu$_{0.595}$Y$_{0.405}$)$_{1-x}$Tb$_x$MnO$_3$ is much larger than that of Eu$_{0.595}$Y$_{0.405}$MnO$_3$\@. The observed $M_a$ is larger than $M_b$ and $M_c$ over the whole temperature region, and magnetic anisotropy develops with decreasing temperature or increasing Tb concentration. Figure~\ref{fig2} shows the temperature dependence of the ferroelectric polarization of (Eu$_{0.595}$Y$_{0.405}$)$_{1-x}$Tb$_x$MnO$_3$ with 0~$\le x \le$~0.2 in a zero magnetic field. With increasing Tb concentration, the $P_c$ phase develops abruptly, and $T_{{\rm FE}c}$ rises slightly. In contrast, the $P_a$ phase is more fragile against Tb doping, and it totally disappears at a Tb-doping level of $x$~=~0.2\@. In $x \geq 0.2$, (Eu$_{0.595}$Y$_{0.405}$)$_{1-x}$Tb$_x$MnO$_3$ shows magnetoelectric behavior similar to that of TbMnO$_3$\@. As we previously reported,\cite{japnoda} this is probably because of the internal magnetic field arising from magnetic coupling between Mn~3$d$ spins and Tb~4$f$ moments. We summarize these results in the magnetoelectric phase diagram shown~in~Fig.~\ref{fig3}(a). 

The effect of Gd substitution contrasts sharply with that of Tb substitution. 
In (Eu,Y)$_{0.9}$Gd$_{0.1}$MnO$_3$, the $P_c$ phase is squeezed into a narrow region of $T$~=~23~K~-~24~K\@, and the magnetic anisotropy is quite small compared to that of (Eu$_{0.595}$Y$_{0.405}$)$_{0.9}$Tb$_{0.1}$MnO$_3$ [Figs.~\ref{fig1}(e)~and~\ref{fig1}(f)]. 
Figure \ref{fig4} shows the temperature dependence of $P_a$ and $P_c$ in (Eu,Y)$_{1-x}$Gd$_x$MnO$_3$ with~0 $\le x \le$~0.2 in a zero magnetic field. As seen from Fig.\ \ref{fig4}, $P_a$ is relatively robust against Gd substitution, and it still remains up to $x$~=~0.69\@. However, $T_{{\rm FE}a}$ shifts to lower temperatures with increasing Gd concentration. In the end material, Y$_{0.31}$Gd$_{0.69}$MnO$_3$\@, $T_{{\rm FE}a}$ is reduced by 8~K\@. On the other hand, $P_c$ is not very robust against Gd-doping, and it is totally extinguished at $x \ge$~0.2\@. The magnetoelectric phase diagram of (Eu,Y)$_{1-x}$Gd$_x$MnO$_3$ with 0~$\le x \le$~0.69 is displayed~in~Fig.~\ref{fig5}.

\section{Discussion}
\begin{figure}[tb]
\begin{center}
\includegraphics[width=0.45 \textwidth, clip]{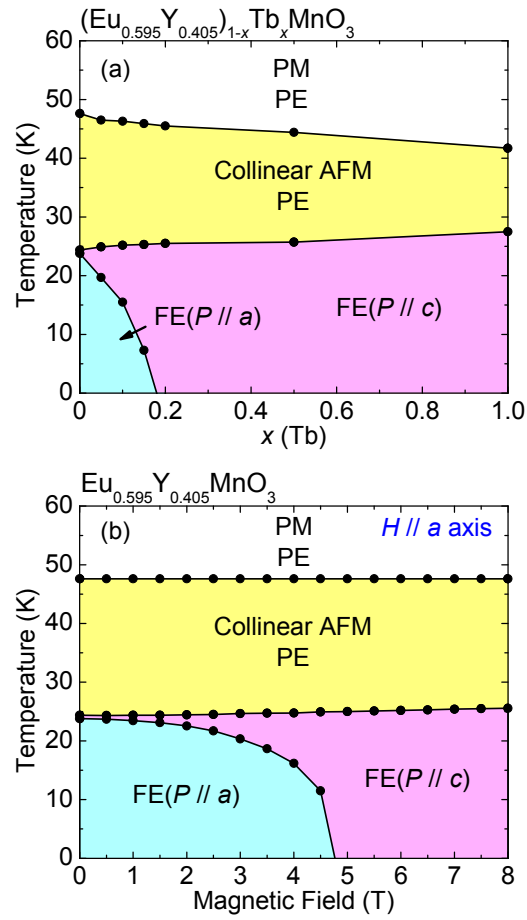}
\end{center}
\caption{(Color online) Magnetoelectric phase diagram of (Eu$_{0.595}$Y$_{0.405}$)$_{1-x}$Tb$_x$MnO$_3$ with 0~$\le x \le$~1 (a), and that of Eu$_{0.595}$Y$_{0.405}$MnO$_3$ as a function of applied magnetic field parallel to the $a$-axis ($H_a$) (b).\cite{japnoda} Abbreviations denote the paramagnetic (PM), paraelectric (PE), antiferromagnetic (AFM), and ferroelectric (FE) phases. Data points are obtained from the measurements of the dielectric constant.}
\label{fig3}
\end{figure}

\begin{figure}[tb]
\begin{center}
\includegraphics[width=0.4 \textwidth, clip]{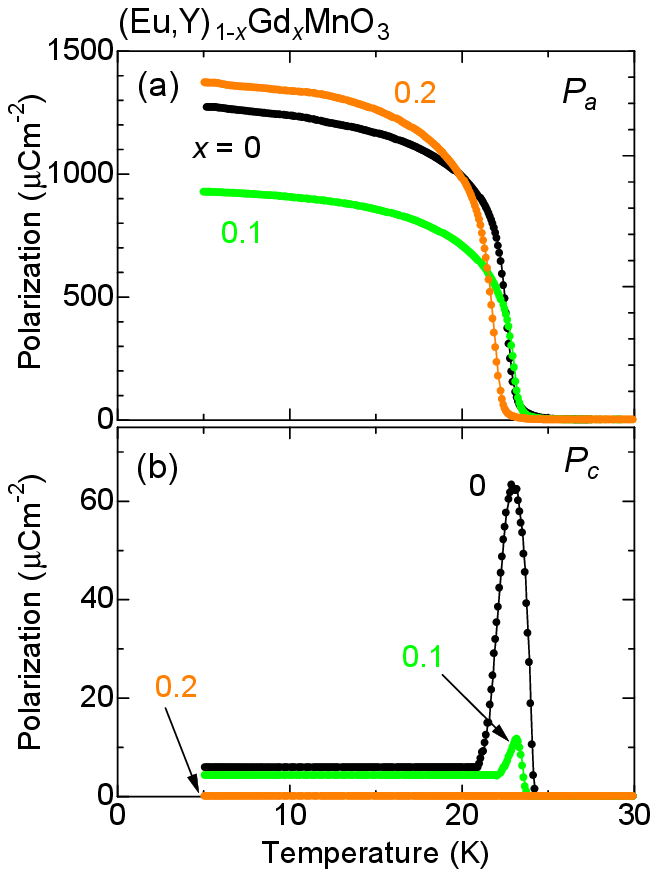}
\end{center}
\caption{(Color online) Temperature dependence of ferroelectric polarization along the $a$ ($P_a$,~upper) and $c$ ($P_c$,~lower) axes of (Eu,Y)$_{1-x}$Gd$_x$MnO$_3$ with 0~$\le x \le$~0.2 in~a~zero~magnetic~field.}
\label{fig4}
\end{figure}

\begin{figure}[tb]
\begin{center}
\includegraphics[width=0.38 \textwidth, clip]{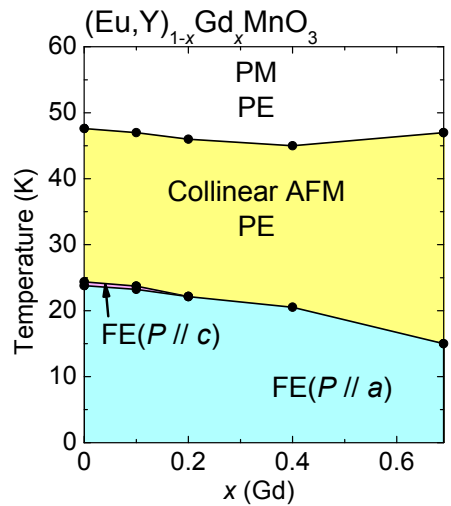}
\end{center}
\caption{(Color online) Magnetoelectric phase diagram of (Eu,Y)$_{1-x}$Gd$_x$MnO$_3$ with 0~$\le x \le$~0.69\@. Abbreviations denote the paramagnetic (PM), paraelectric (PE), antiferromagnetic (AFM), and ferroelectric (FE) phases. Data points are obtained from the measurements of the dielectric constant.}
\label{fig5}
\end{figure}

First, let us consider the effect of Tb substitution. As shown in the magnetoelectric phase diagram of (Eu$_{0.595}$Y$_{0.405}$)$_{1-x}$Tb$_x$MnO$_3$ (0~$\le x \le$~1) [Fig.~\ref{fig3}(a)], with increasing Tb concentration, $T_{{\rm FE}a}$ falls sharply, and the $P_a$ phase completely disappears at $x = 0.2$\@. 
On the other hand, the $P_c$ phase exists for all values of $x$\@, and $T_{{\rm FE}c}$ rises slightly with increasing Tb concentration. We display in Fig.~\ref{fig3}(b) the magnetoelectric phase diagram of Eu$_{0.595}$Y$_{0.405}$MnO$_3$ as a function of applied magnetic field parallel to the $a$-axis ($H_a$) for comparison with the magnetoelectric phase diagram of (Eu$_{0.595}$Y$_{0.405}$)$_{1-x}$Tb$_x$MnO$_3$.\cite{japnoda} In Eu$_{0.595}$Y$_{0.405}$MnO$_3$, with increasing~$H_a$\@, $T_{{\rm FE}a}$ decreases, and $P_a$ is extinguished in $H_a$ $\geq$~5~T\@. 
The $P_c$ phase exists over the whole $H_a$ region, and $T_{{\rm FE}c}$ is almost invariable against~$H_a$\@.
This is because the application of $H_a$ destabilizes the $ab$ spiral that induces the $P_a$ phase. Namely, $H_a$ perpendicular to the $bc$ spiral plane induces a conical state and gains Zeeman energy.
As clearly seen in Figs.~\ref{fig3}(a)~and~\ref{fig3}(b), the magnetoelectric phase diagram of (Eu$_{0.595}$Y$_{0.405}$)$_{1-x}$Tb$_x$MnO$_3$ is qualitatively similar to that of Eu$_{0.595}$Y$_{0.405}$MnO$_3$ as a function of~$H_a$\@. 
From this similarity, one can conclude that the effect that Tb substitution has on the Mn$^{3+}$ sublattice as similar to the effect the application of $H_a$ has. 
The enhancement of $M_a$ induced by Tb substitution strongly supports the above conclusion [Fig.~\ref{fig1}(d)]. 
The $P_a$ phase of Eu$_{0.595}$Y$_{0.405}$MnO$_3$ is destroyed by an external magnetic field of $H_a$~=~5~T\@, while that of (Eu$_{0.595}$Y$_{0.405}$)$_{1-x}$Tb$_x$MnO$_3$ disappears at $x = 0.2$\@. 
Thus, the internal magnetic field of (Eu$_{0.595}$Y$_{0.405}$)$_{0.8}$Tb$_{0.2}$MnO$_3$ can be estimated to be 5~T\@. 

Next, we discuss the effect of Gd substitution. As Figs.\ \ref{fig3} and \ref{fig5} show, the magnetoelectric phase diagram of (Eu,Y)$_{1-x}$Gd$_x$MnO$_3$ (0~$\le x \le$~0.69) is quite different from that of (Eu$_{0.595}$Y$_{0.405}$)$_{1-x}$Tb$_x$MnO$_3$\@. 
With increasing Gd concentration, both $T_{{\rm FE}c}$ and $T_{{\rm FE}a}$ decrease, and the $P_c$ phase cannot be observed at $x \geq 0.2$. 
This difference originates from the difference in the characteristics of the 4$f$ moments of Gd$^{3+}$ and Tb$^{3+}$\@, which greatly affect the magnetic anisotropy of $R$MnO$_3$\@. 
The magnetic anisotropy of (Eu,Y)$_{1-x}$Gd$_x$MnO$_3$ is much smaller than that of (Eu$_{0.595}$Y$_{0.405}$)$_{1-x}$Tb$_x$MnO$_3$ [Figs.~\ref{fig1}(d)~and~\ref{fig1}(f)], because of the isotropic 4$f$ moment of Gd$^{3+}$\@. 
The nearly isotropic magnetization of (Eu,Y)$_{1-x}$Gd$_x$MnO$_3$ indicates that magnetic coupling between Gd 4$f$ moments and Mn 3$d$ spins does not enhance the magnetization in a particular direction, which is in contrast to the case of (Eu$_{0.595}$Y$_{0.405}$)$_{1-x}$Tb$_x$MnO$_3$. Therefore, it does not act as an internal magnetic field.

Another remarkable difference between the two systems is that both ferroelectric transition temperatures ($T_{{\rm FE}a}$~and~$T_{{\rm FE}c}$) decrease in (Eu,Y)$_{1-x}$Gd$_x$MnO$_3$ but not in (Eu$_{0.595}$Y$_{0.405}$)$_{1-x}$Tb$_x$MnO$_3$\@. The decrease in the ferroelectric transition temperatures is probably due to $R$-site randomness arising from solid solution of the $R$-site ions and/or magnetic randomness arising from random distribution of Gd$^{3+}$ or Tb$^{3+}$ in $R$ sites. 
Generally, the $R$-site randomness can be measured by the variance $\sigma^2 = \Sigma(g_ir_i^2-r_R^2)$, where $g_i$, $r_i$ ($i = 1 \sim 3$), and $r_R$ are the fractional occupancy, the ionic radius of the $i$-th ion, and the average ionic radius of $R$ sites, respectively.\cite{Tokura,Att} $r_R$ is equal to the ionic radius of Tb$^{3+}$.  $\sigma^2$, i.e., the $R$-site randomness of both systems, becomes smaller with increasing Gd or Tb concentration. If the $R$-site randomness effect were dominant, Gd or Tb substitution could raise the ferroelectric transition temperatures. However, Gd substitution actually reduces these transition temperatures. 
Spiral antiferromagnetic order arises from subtle balance between the NN ferromagnetic and the next NN antiferromagnetic interactions so that it would be rather sensitive to magnetic randomness. As a result, Gd substitution reduces $T_{{\rm FE}c}$ and $T_{{\rm FE}a}$. On the contrary, in the Tb-substituted sample, the internal magnetic field effect, which stabilizes the $P_c$ phase, is larger than magnetic randomness effect. Consequently, $T_{{\rm FE}c}$ of the Tb-substituted sample rises slightly.

\section{Conclusion}
In this study, we have established detailed magnetoelectric phase diagrams of (Eu$_{0.595}$Y$_{0.405}$)$_{1-x}$Tb$_x$MnO$_3$ and (Eu,Y)$_{1-x}$Gd$_x$MnO$_3$ to clarify the effect of rare earth 4$f$ moments on the magnetoelectric properties; the average $R$-site ionic radii of the two systems equal that of Tb$^{3+}$.
The parent compound Eu$_{0.595}$Y$_{0.405}$MnO$_3$ without 4$f$ moments undergoes the $P_c$ transition at 25~K\@, and then the $P_a$ transition at 23~K\@. 
Tb substitution has the similar effect on the Mn$^{3+}$ sublattice as applying $H_a$ (we refer to this effective magnetic field as the internal magnetic field). 
As a result, the magnetoelectric phase diagram of (Eu$_{0.595}$Y$_{0.405}$)$_{1-x}$Tb$_x$MnO$_3$ is qualitatively similar to that of Eu$_{0.595}$Y$_{0.405}$MnO$_3$ as a function of $H_a$\@.
On the other hand, in (Eu,Y)$_{1-x}$Gd$_x$MnO$_3$, the internal magnetic field is ineffective because of an isotropic 4$f$ moment of Gd$^{3+}$. With increasing Gd concentration, $T_{{\rm FE}a}$ and $T_{{\rm FE}c}$ are decreasing, and the $P_c$ phase is totally extinguished in $x \ge$~0.2\@.
Probably, magnetic randomness coming from random distribution of Gd$^{3+}$ in $R$-site is the main cause of the decrease in the ferroelectric transition temperetures. 
The present results reveal the significant contribution of 4$f$ moments to the magnetoelectric properties of $R$MnO$_3$\@, and in addition, provide significant information to control the magnetoelectric properties using rare earth 4$f$ moments. 

\section*{Acknowledgment}
This work was partly supported by Grant-in-Aid for Scientific Research (C) from the Japan Society for the Promotion of Science (JSPS) and for JSPS Fellows.

\newpage

\end{document}